\documentclass[superscriptaddress,aps,twocolumn,pra,showpacs,floatfix]{revtex4-1}
\usepackage{epsfig,amsbsy,bm,amssymb,amsmath}
\usepackage{graphicx}
\usepackage{graphics}

\begin{document}
\bibliographystyle{apsrev}

\renewcommand{\k}{{\bbox k}}
\newcommand{\hs}{\hspace*}
\newcommand{\vs}{\vspace*}
\newcommand{\np}{\newpage}
\newcommand{\cx}{C$_{60}$~}
\newcommand{\eref}[1] {(\ref{#1})}
\newcommand{\Eref}[1] {Eq.~(\ref{#1})}
\newcommand{\Fref}[1] {Fig. \ref{#1}}
\newcommand{\ra}{\rangle}
\newcommand{\la}{\langle}
\newcommand{\nn}{\nonumber}
\newcommand{\be}{\begin{equation}}
\newcommand{\ee}{\end{equation}}
\newcommand{\br}{\begin{eqnarray*}}
\newcommand{\er}{\end{eqnarray*}}
\newcommand{\ba}{\begin{eqnarray}}
\newcommand{\ea}{\end{eqnarray}}
\newcommand{\bp}{\begin{minipage}}
\newcommand{\ep}{\end{minipage}}
\newcommand{\ds}{\displaystyle}
\newcommand{\bs}{\bigskip}

\title{Attosecond time delay in the photoionization of Mn in the region of the $3p \to 3d$ giant resonance}

\author{V. K. Dolmatov}
\affiliation{Department of Physics and Earth Science, University of North
Alabama, Florence, AL 35632, USA}
\author{A. S. Kheifets}
\affiliation{Research School of Physics and Engineering, The
Australian National University, Canberra ACT 0200, Australia}
\author{P. C. Deshmukh}
\affiliation{Department of Physics, Indian Institute of Technology
  Madras, Chennai-600036, India}
\author{S. T. Manson}
\affiliation{Department of Physics and Astronomy, Georgia State
University, Atlanta, 30303, U.S.A.}

\date{\today}

\begin{abstract}
The initial insight into time delay in Mn photoionization in the region
of the $3p \to 3d$ giant autoionization resonance is gained in the framework
of the ``spin-polarized'' random phase approximation with exchange.
The dramatic effect of the giant
autoionization resonance on time delay of photoemission from the
$3d$ and $4s$ valence subshells of the Mn atom is unraveled.
Strong sensitivity of the time delay of the $4s$ photoemission to the
final-state term of the ion-remainder [${\rm Mn^{+}}(4s^{1},$$^{5}S)$
vs.~${\rm Mn^{+}}(4s^{1},$$^{7}S)$] is discovered. It is shown that photoionization time delay in the autoionizing resonance region
is explicitly associated with the resonance lifetime, which can, thus, be directly measured
in attosecond time delay experiments. Similar features are expected to emerge in photoionization time delays of other transition-metal and rare-earth atoms
with half-filed subshells that possess giant autoionization resonances as well.
\end{abstract}

\maketitle

\section{Introduction}
Atomic photoionization time delay is characterized by a slight temporal
offset in the release of the photoelectron wave-packet upon
absorption of a short electromagnetic pulse by an atom.
Since the first experimental and theoretical demonstration that atomic
photoionization time delay can be measured using a single attosecond
pulse (SAP) \cite{M.Schultze06252010}, it has become an active topic
of investigation. A series of experiments was conducted recently using
attosecond pulse trains (ATP) \cite{PhysRevLett.106.143002,PhysRevA.85.053424,Guenot2014,DiMauro2014,PhysRevLett.112.153001}.
Thus produced experimental results were analyzed using several
theoretical models of various degrees of sophistication
\cite{PhysRevLett.105.233002,
PhysRevA.84.061404,
0953-4075-44-8-081001,
PhysRevA.86.061402,Dahlstroem2012b,PhysRevA.87.063404,Dahlstrom2012,
PhysRevA.89.033417,PhysRevA.90.053406}.
At the same time, there exists a large body of literature on
photoemission time delay in condensed matter but reviewing this
literature is outside the framework of the present article.

In essence, photoionization time delay is a direct generalization of
the concept of time delay developed by \citet{Eisenbud1948} and
\citet{PhysRev.98.145} for electron scattering and applied recently to
atomic photoionization in Ref.~\cite{M.Schultze06252010}. Normally, the delay is small, of the order of
tens to hundreds of attoseconds ($\rm 1~as=10^{-18}~s$). Experimental
observation of this phenomenon allows one to capture electron motion in
atoms, molecules and solids on its natural, attosecond time scale. In
turn, unique experimental accomplishments provide the impetus for
advanced theoretical studies of the photoionization time delay
phenomenon as well.

To date, to the authors' best knowledge, photoionization time delay
has only been studied in closed shell systems like noble gas atoms.
However, there is another interesting group of atoms, the
transition-metal atoms, where \textit{photoionization} time delay has
not been studied at all yet; it was only addressed, briefly, for elastic electron
scattering off Mn \cite{PhysRevA.88.042706}. Meanwhile, time
delay in the photoemission spectra of transition-metal atoms presents an
especially interesting case. Owing to the open-shell nature of the valence
$nd^{q<10}$-subshells of these atoms ($n=3$ for iron-group atoms, like
the Mn atom), their photoionization spectra are dominated by the $np
\to nd$ giant autoionization resonance which subsequently
autoionizes into primarily $nd \to f, p$ channels. The $np
\to nd$ giant resonance was originally detected experimentally in the
$3p$-photoabsorption spectrum of Mn by \citet{Connerade1976}. Later,
it was experimentally and theoretically studied not only in Mn but in
other transition-metal atoms and their ions as well (see review
papers by \citet{0034-4885-55-7-002} and \citet{0953-4075-39-5-R01},
as well as references cited below in this paper).

It is the ultimate aim of the present study to get insight into the impact
of the $3p \to 3d$ giant autoionization resonance in the Mn([Ar]$3d^{5}4s^2,^{6}S)$
atom on time delays in photoionization of the $3d$ and $4s$ valence subshells of the atom.

The effect of resonances on measuring and interpreting atomic photoemission
time delay has been studied previously. In Ref.~\cite{Athens} doubly
excited states of Ne were scrutinized to explain the discrepancy
between experiment and theory reported in
Ref.~\cite{M.Schultze06252010}. Similar attempts were made to
reconcile the theory and experiment for photoemission time delay near
the $3s$ Cooper minimum in Ar \cite{PhysRevA.87.023420}. However,
neither of these attempts was successful. In Ne, the resonances proved
to be too narrow to have any effect on the measured time delay. In Ar,
the latest and most accurate set of time delay results
\cite{2014arXiv1407.6623S} was found in disagreement with the theoretical
predictions of \cite{PhysRevA.87.023420}.

There are reasons for choosing Mn for this study. First,
$3d$-photoionization of neutral Mn in the region of the $3p \to
3d$ resonance was studied extensively experimentally
\cite{0034-4885-55-7-002,0953-4075-39-5-R01,PhysRevA.30.1316,Whitfield94,0953-4075-45-22-225204}.
Thus, Mn is well disposed for experimental photoionization
measurements, and there is a reliable experimental information to
assess the quality of corresponding theoretical calculations. Second,
Mn is not just an open-shell atom but a half-filled shell atom.  This
simplifies its theoretical study significantly. In particular, one can
employ a multielectron ``spin-polarized'' random phase approximation
with exchange (SPRPAE) \cite{JETP1983,0953-4075-26-8-010,AC97}
designed especially to describe photoionization of half-filled shell
atoms. Finally, SPRPAE has been successfully used for the study of the
$3p \to 3d$ giant resonance in $3d$ \cite{JETP1983} and $4s$
\cite{JPB88,JPB90} photoionization of Mn, and good quantitative agreement with
experiment \cite{0034-4885-55-7-002,PhysRevA.30.1316,Whitfield94} and
many-body perturbation theory (MBPT)
\cite{0022-3700-16-9-004}, was achieved. Thus, SPRPAE is a convenient
theoretical method for gaining the initial insight into time delays in Mn photoionization.
It is, therefore, chosen as the theoretical  tool for the present study as well.

Atomic units are used throughout the paper unless specified otherwise.

\section{Review of theory}

A convenient starting point to account for the structure of a
half-filled shell atom is provided by the spin-polarized Hartree-Fock (SPHF)
approximation developed by  \citet{slater1974self}.
SPHF accounts for the fact that spins of all electrons in a
half-filled subshell of the atom (e.g., in the $\rm 3d^{5}$ subshell
of Mn) are aligned, in accordance with the Hund's rule, say, all are
pointing upward. This results in splitting of a closed
${n\ell}^{2(2\ell+1)}$ subshell in the atom into two half-filled
subshells of opposite spin orientations, ${n\ell}^{2\ell+1}$$\uparrow$
and ${n\ell}^{2\ell+1}$$\downarrow$.  This is due to the presence
of the exchange interaction between $n\ell$$\uparrow$ electrons with
only spin-up electrons of a spin-unpaired half-filled subshell of the
atom (e.g., the $ 3d^{5}$$\uparrow$ subshell in the Mn atom), but
the absence of such interaction for $n\ell$$\downarrow$
electrons. Therefore, atoms with half-filled subshells can be treated
as having only occupied subshells, filled in only by either one or the other kind of electrons,
named ``\textit{up}''- or ``\textit{down}''-electrons depending on their
spin orientations, $\uparrow$ and $\downarrow$, respectively
\cite{JETP1983,slater1974self}. Their binding energies
$\epsilon_{n\ell\uparrow(\downarrow)}$ and wave functions
$P_{\epsilon\uparrow(\downarrow)}(r)$ differ from each other, as is
clear from the discussion above. They are solutions of the
corresponding SPHF equations which differ from the ordinary
Hartree-Fock equations by accounting for exchange interactions only
between electrons with the same spin-orientation
\cite{AC97,slater1974self}.
For the Mn atom, which is the atom of interest of the present paper,
the SPHF configuration is
[Ar]${3p}^{3}$$\uparrow$${3p}^{3}$$\downarrow$${3d}^{5}$$\uparrow$${4\rm
s}^{1}$$\uparrow$$4s^{1}$$\downarrow$ ($^{6}S_{5/2}$).
The removal of a $3d$$\uparrow$-electron produces the ion-remainder
Mn$^{+}$($3d^{4}$$\uparrow$ $4s^{1}$$\uparrow$ 4s$\downarrow$, $^{5}D_{4})$.  A removal of a spin-up
$4s$$\uparrow$ or spin-down $4s$$\downarrow$ electron from Mn results in
ion-remainders having different terms, the Mn$^{+}$($3d^{5}$$\uparrow$ $4s^{1}$$\downarrow$, $^{5}S_{2})$ or
Mn$^{+}$($3d^{5}$$\uparrow$ $4s^{1}$$\uparrow$, $^{7}S_{3})$ ions, respectively. This
makes the photoionization process spin-dependent, or, in other words,
term-dependent. In the present paper, ``term dependence'' and ``spin
dependence'' are used interchangeably.

\begin{figure}
\includegraphics[width=7cm]{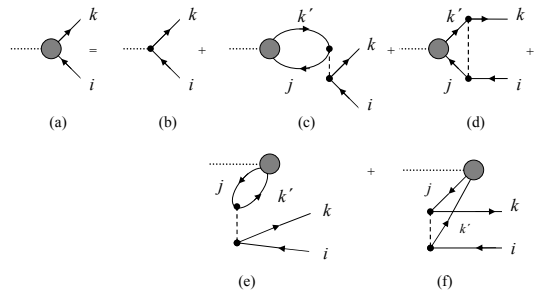}
\caption{Feynman
diagrammatic representation of the SPRPAE (RPAE) equation for the
photoionization amplitude $\la k|\hat{D}|i\ra $ of the \textit{i}'th subshell
into the \textit{k}'th final state \cite{AC97}. Here, the time axis is directed from the left to right,
the lines with
arrows to the left (right) correspond to holes (electrons) in the atom, a
dotted line represents an incoming photon, a dashed line represents the
Coulomb interaction $V(r)$ between charged particles, and a shaded circle marks
 the effective operator $\hat{D}$ for the photon-atom interaction which accounts for electron
correlation in the atom.}
\label{fig1}
\end{figure}

The multielectron SPRPAE~\cite{JETP1983,0953-4075-26-8-010,AC97} utilizes
SPHF as the zero-order independent-particle basis - the vacuum
state. This is because the \textit{spin-up}- and
\textit{spin-down}-subshells of the atom can be regarded as completely
filled. Therefore, the well-developed random phase approximation with
exchange (RPAE) for closed shell atoms \cite{AC97} can be easily
generalized to the case of half-filled shell atoms.  Similar to RPAE,
the SPRPAE equation for a photoionization amplitude $\la k|\hat{D}|i\ra
\equiv D_{ki}$ of the \textit{i}'th subshell of an atom into a
continuous state $k$ is depicted graphically in Fig.~\ref{fig1}.
There, diagrams (c)--(f) represent SPRPAE (RPAE) corrections to the
HF photoionization amplitude $\la k|\hat{d}|i\ra \equiv d_{ik}$
[diagram (b)].  Diagrams (d) and (f) account for the exchange interaction
in the atom, thus being called the exchange diagrams. In SPRPAE, the
contribution of the exchange diagrams (d) and (f) to the photoionization amplitude
is discarded from the equation whenever the corresponding
intermediate-state electron-hole pair ``$j-k'$'' and the final-state
electron-hole pair ``$i-k$'' have opposite spin orientations.   The exchange diagrams (d) and (f), in fact,
represent an infinite sum over all orders of perturbation theory in
the inter-electron interaction. Therefore, the presence of
certain series of exchange diagrams in the SPRPAE equation for photoionization of
an electron with one spin-polarization (e.g., a spin-up electron) but the absence of this series of
exchange diagrams in
the corresponding equation for photoionization of an electron with an opposite spin-orientation (spin-down electron, in
this example) matters a lot. This
enhances the photoionization term-dependence of a half-filled shell atom considerably
compared to SPHF results. In fact, such term dependence was
found to be dramatic not only for dipole photoionization of the
outermost $ns$-subshells in half-filled shell atoms
\cite{JPB88,JPB90,DolmJPB96}, but for nondipole photoionization as
well \cite{DolmPRA06}.

We now briefly outline the key points of the photoionization time
delay concept. In the spirit of the Eisenburg-Wigner theory for time
delay in electron scattering \cite{Eisenbud1948,PhysRev.98.145}, time
delay in the photoionization of a $n_{i}l_{i}$ subshell of the atom is
determined by a derivative of the phase $\varphi(E)$ of corresponding
photoionization amplitude $T_{n_{i}l_{i}}=
\left|T_{n_{i}l_{i}}\right|e^{i\varphi(E)}$ \cite{PhysRev.118.349}.
Correspondingly,
\ba
 \varphi(E) = \arg[T_{n_{i}l_{i}}(E)], \quad \tau_{n_{i}l_{i}}= \frac{d\varphi(E)}{dE}.
\label{phase}
\ea
For a photoionization amplitude $T_{n_{i}l_{i}}$ of a $n_{i}l_{i}$-state which accounts for both
$n_{i}l_{i} \to \epsilon(l_{i} \pm 1)$ dipole transitions, one has
\cite{PhysRevA.90.053406}:
\ba
\label{amplitude}
T_{n_il_i}(E)&\propto&
\sum_{l=l_i\pm1\atop m=m_i}
e^{i\delta_l}i^{-l}
Y_{lm}(\hat{\bm k})\,
(-1)^m
\left(\begin{array}{rrr}
l&1&l_i\\
-m&0&m_i\\
\end{array}\right)
\nn\\&&\hs{2cm}\times \
 \la El\|D\|n_il_i \ra.
\ea
Here, $\hat {\bm k}$ is a unit vector in the direction of the
photoelectron momentum $\bm k$, $\delta_{l}(E)$ is the phase shift of
the $l$th outgoing photoelectron wave, and $\la El\|D\|n_il_i \ra$ is
the reduced dipole matrix element
which is the solution of the RPAE (or SPRPAE, in our work)
equation, Fig.~\ref{fig1}. Since the  matrix element $T_{n_{i}l_{i}}$ defined by Eq.~(\ref{amplitude})
depends on the photoelectron emission angles, it will be referred to, when the emphasis is needed, as the angle-dependent
matrix element.
In the present work, $T_{n_{i}l_{i}}(E)$ is evaluated in the forward direction
${\bm k}\|\hat z$, which is usually the case in the attosecond time delay
measurements; this is of importance because the time delay, in
general, has an angular dependence \cite{0953-4075-48-2-025602,Dahlstrom2014}.

Summarizing the review of theory, the Eisenbud-Wigner time delay
\eref{phase} is defined by the complex phase of the stationary
ionization amplitude corresponding to a given energy of the ionizing
field $\omega$. However, in SAP or ATP measurements, time delay is
determined by a combination of the ionizing field and the streaking
probe that is used. This introduces the so-called Coulomb-laser
coupling corrections \cite{0953-4075-44-8-081001} (SAP) or
continuum-continuum corrections \cite{Dahlstroem2012b} (ATP). The
latter reference also demonstrated equivalence of the corrections in
the SAP and ATP measurements.  In non-resonant photoionization, these
corrections should be added to the Eisenbud-Wigner time delay to
account for the experimentally measured time delay.

 In resonant photoionization, the effect of the measurement on the Eisenbud-Wigner time
 delay is much more complicated. Indeed, the SAP measurement involves an attosecond pulse
 of a considerable bandwidth which is required to be much larger than the photon energy of
 the probing pulse  $\Omega\simeq1.5$~eV \cite{Dahlstroem2012b}. Such wide spectral bandwidth could, at least partially,
 ``mask'' an autoionization resonance of comparable width which is manifested in the present
 Mn $3p\to3d$  case, so that it might be difficult to employ the SAP technique to verify the
 made predictions. On the other hand, although the ATP technique is free from the bandwidth
  limitations, the usual two-frequency modulation of the two-photon interference is
  strongly distorted by the presence of an autoionizing resonance \cite{PhysRevLett.113.263001}. Hence a more involved
  analysis is required that cannot be reduced to adding a unified set of the continuum-continuum corrections.
   Recently, an analysis is provided by a general analytical model that accounts for the effect
   of both intermediate and final resonances on two-photon processes \cite{PhysRevLett.113.263001}.  With the helium atom
   as the case study, this analysis yields the same result as a solution of the all-dimensional
    time-dependent Schr\"{o}dinger equation. Therefore, the ATP technique looks more suitable than
    SAP to address a resonance problem, although it is not entirely clear how it could account
    for the interchannel interference.
\section{Results and discussion}
\subsection{Mn $3d$-photoionization}

The SPRPAE calculations of Mn $3d$-photoionization in the region of
the $3p$$\downarrow \to 3d$$\downarrow$ giant autoionization resonance
were performed including interchannel coupling among four
$3d$$\uparrow \to f$$\uparrow$, $3d$$\uparrow \to p$$\uparrow$,
$3p$$\downarrow \to d$$\downarrow$, and $3p$$\downarrow \to
s$$\downarrow$ transitions. Interchannel coupling with other
transitions was ignored as negligible. Next, calculated SPHF values
for the ionization potentials $I_{3d\uparrow} \approx 17.4$ eV and
$I_{3p\downarrow} \approx 60.7$ eV, as well as binding energies of
discrete excitations were used in this calculation. This is because the use
of HF (SPHF) ionization potentials is conceptually consistent with
RPAE (SPRPAE) theory. Moreover, earlier \cite{JETP1983}, the use of
SPHF ionization thresholds in the calculated SPRPAE
$3d$-photoionization cross section of Mn in the $3p \to 3d$ resonance
region was shown to result in a good agreement between theory and
experiment.

\begin{figure}
\includegraphics[width=7cm]{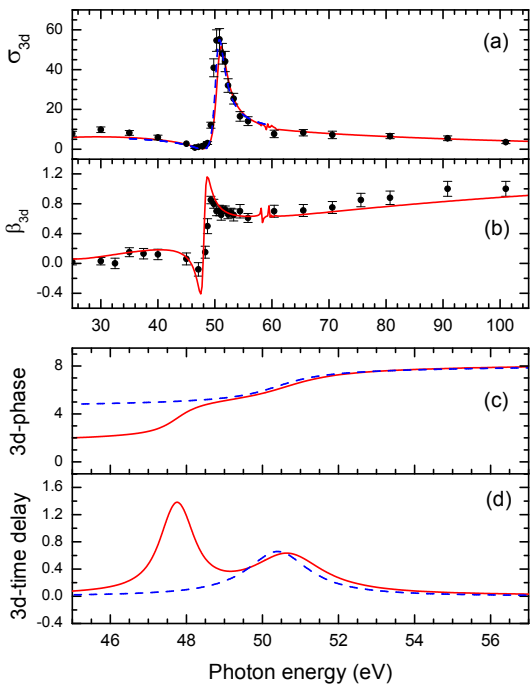}
\caption{(Color online)
(a): solid line - present calculated SPRPAE data for the Mn
$3d$$\uparrow$-photoionization cross-sections $\sigma_{3d}$ (in units of Mb);
dots - experimental data on an absolute scale taken form Table $3$ of Ref.~\cite{PhysRevA.30.1316};
dashed line - SPRPAE  $\sigma_{3d}$ calculated previously \cite{JETP1983} on the basis of the Fano
autoionization formalism in the $3p$$\downarrow$ $\to$ $3d$$\downarrow$ giant resonance region.
(b): solid line - present calculated SPRPAE angular-asymmetry
parameter $\beta_{3d}$; dots - experimental data from Table $3$ of Ref.~\cite{PhysRevA.30.1316}.
(Note, former SPRPAE results for  $\beta_{3d}$ \cite{JETP1983}, obtained on the basis of the Fano
formalism in the $3p$$\downarrow$ $\to$ $3d$$\downarrow$ giant resonance region, practically coincide
with the present data and not plotted in this figure).
(c): solid line - calculated SPRPAE phase $\varphi_{3d}(\omega)$ (in units of
radians) of the $3d$$\uparrow$-photoionization amplitude $T_{3d}$, Eq.(\ref{amplitude});
dashed line - the same as above but calculated with the help of the single-channel
Fano formula, Eq.~(\ref{Fano}).
(d): solid line - calculated SPRPAE time
delay $\tau_{3d}(\omega)$ (in units of femtoseconds); dashed line - $\tau_{3d}(\omega)$
calculated with the help of the single-channel Fano formula, Eq.~(\ref{tauf}).}
\label{fig2}
\end{figure}

The present calculated SPRPAE results for the
$3d$$\uparrow$-photoionization cross section $\sigma_{3d}(\omega)$ and
$3d$-angular-asymmetry parameter $\beta_{3d}$ (refer, e.g.,
to Ref.~\cite{AC97} for the equations for $\sigma_{nl}$ and $\beta_{nl}$)
are plotted in Fig.~\ref{fig2}(a) and (b) between $25$ and $105$ eV along with corresponding
experimental data from Ref.~\cite{PhysRevA.30.1316}. One can see reasonable agreement not
only between theory and experiment for $\sigma_{3d}(\omega)$ but for $\beta_{3d}$ as well.
The angular-asymmetry parameter $\beta_{3d}$ depends not only on the absolute values of the
photoionization amplitudes but also on phase shifts of the amplitudes. Therefore, reasonable agreement
between the present calculated SPRPAE data and experiment for $\beta_{3d}$ is indicative of reasonably
correctly calculated phases of the photoionization matrix elements and, thus, $3d$-time delay $\tau_{3d}$ as well.
The calculated SPRPAE phase $\varphi_{3d}(\omega)$ and
time delay $\tau_{3d}(\omega)$ in the region of the
$3p$$\downarrow$~$\to$~$3d$$\downarrow$ giant resonance are depicted  in
Fig.~\ref{fig2}c and Fig.~\ref{fig2}d, respectively.

Note also how significantly the giant resonance impacts both $\varphi_{3d}(\omega)$ and
$\tau_{3d}(\omega)$, compared to the region away from the
resonance. Specifically, the giant resonance enhances the time delay,
$\tau_{3d}(\omega)$, by more than one order of magnitude compared to
its nonresonance value, both at  $\omega \approx 48$ and $50.5$ eV.
It is important to note that the latter enhancement of
$\tau_{3d}(\omega)$ occurs in the photon energy region where
the cross section is large, $\sigma_{3d} \approx 55$ Mb.
This should facilitate greatly its experimental observation.

It is also instructive to make a simple evaluation of a
photoionization amplitude $T_{3d}$, its phase $\varphi_{3d}$, and
time delay $\tau_{3d}$ in the $3p$$\downarrow \to 3d$$\downarrow$ giant resonance region within
the framework of the Fano theory \cite{PhysRev.124.1866}.
A single-channel, single-resonance parametric
expressions for the generally dominant matrix element $T_{3d \to f}(\omega) \equiv T$, Eq.~(\ref{amplitude}), and the photoionization cross section
$\sigma_{3d \to f}(\omega) \equiv \sigma$ read
\be
\label{Fano}
T(\omega) =T_{0} {q+\epsilon\over i+\epsilon}
 \ , \ \
\sigma(\omega)  = \sigma_0 {(q+\epsilon)^2\over 1 +\epsilon^2}
 \ , \ \
\epsilon = {\omega-\omega_{\rm r}\over \gamma/2}.
\ee
Here, $T_{0}$  and $\sigma_{0}$ are, respectively, the $3d$$\uparrow \to f$$\uparrow$ photoionization amplitude and cross section calculated without accounting for
the $3p$$\downarrow \to 3d$$\downarrow$ resonance transition, $\omega_{\rm r}$ is the resonance energy,
$q$ is the profile index (shape parameter),
and $\gamma$ is the resonance width. In Mn, calculated SPRPAE $\gamma \approx 2$ eV, $\omega_{\rm r}\approx 50.4$ eV, $q \approx 2.5$, and $\sigma_{0} \approx 7.6$ Mb \cite{JETP1983}.

With the help of Eqs.~(\ref{phase}) and (\ref{Fano}) one finds
that the time delay $\tau_{3d \to f}$ in a single-resonance, single-channel photoionization Fano formalism is determined as follows:
\begin{eqnarray}
\tau_{3d \to f}(\omega)= \frac{2}{\gamma}\frac{1}{1+\epsilon^2}.
\label{tauf}
\end{eqnarray}
It follows from Eq.~(\ref{tauf}) that single-channel, single-resonance time delay has only one maximum which emerges at $\epsilon =0$, i.e., at the resonance energy $\omega=\omega_{r}$.  Moreover, one readily finds
from Eq.~(\ref{tauf}) that the resonance width $\gamma = 2/\tau_{\rm max}$. The important implication is that the resonance width $\gamma$ at the half-maximum of the photoionization cross section
is explicitly defined by $\tau_{\rm max}$.
This opens a unique
possibility for measuring the autoionization resonance lifetime directly in attosecond
time delay experiments, similar to the corresponding direct measurement of
the Kr $3p$-vacancy Auger decay lifetime \cite{Drescher2002}. Furthermore, it is clear from the above that time delay in the autoionization resonance energy region
reaches its maximum where the photoionization cross section
$\sigma_{nl} =q^2\sigma_0$. One then concludes that the photoionization time delay is easier to
measure in regions of autoionization resonances having large profile indices $q$.

The cross section $\sigma_{3d \to f}$, the phase $\varphi_{3d \to f}$ and
the time delay $\tau_{3d \to f}$ calculated within the framework of the Fano
formalism, discussed above, were depicted in \Fref{fig2}.
 They are in a close
agreement with the  calculated SPRPAE data except for a region near $48$ eV. There, the calculated SPRPAE time delay $\tau_{3d}$ has a strong
resonance structure where $\tau_{3d}$ steeply rises to more than $1.5$~fs below to $48$ eV.
This resonance is then followed by a broader resonance in $\tau_{3d}$ at higher energies where $\tau_{3d}$ reaches about $0.6$~fs near the maximum of $\sigma_{3d}$ at $\omega \approx 50.5$ eV;
the latter is in agreement with the Fano formalism.

\begin{figure}
\includegraphics[width=7cm]{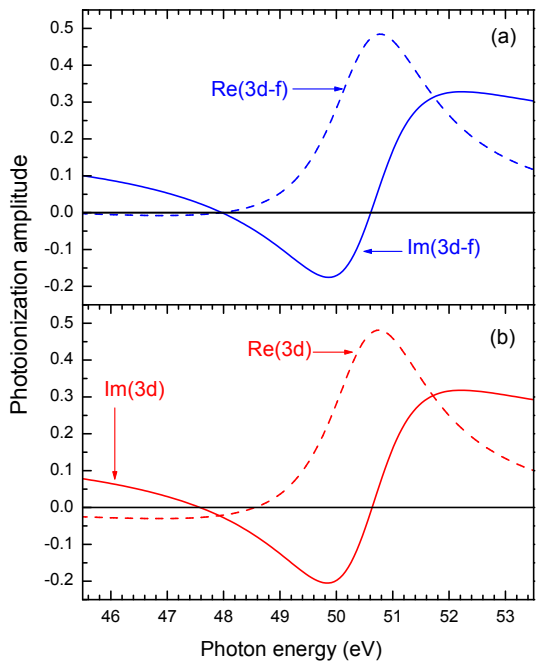}
\caption{(Color online) (a) SPRPAE real ${\rm Re}T_{3d \to f}$  and imaginary ${\rm Im}T_{3d \to f}$ parts (in atomic units) of the dominant $T_{3d \to f}$-photoionization amplitude, Eq.~(\ref{amplitude}),
calculated in the region of the giant $3p$$\downarrow$ $\to$ $3d$$\downarrow$ resonance.
(b) SPRPAE real
${\rm Re}T_{3d}$ and imaginary ${\rm Im}T_{3d}$ parts (in atomic units) of the total amplitude $T_{3d}$,  calculated as the weighted sum of the two
matrix elements $\la Ef\|D\|3d \ra$ and $\la Ep\|D\|3d \ra$, Eq.~(\ref{amplitude}), in the region of the giant $3p$$\downarrow$ $\to$ $3d$$\downarrow$ resonance.}
\label{fig3}
\end{figure}

In order to get insight into the reason for the emergence of the strong $48$-eV resonance in $\tau_{3d}$ but the absence of such in $\tau_{3d \to f}$, we plot in Fig.~\ref{fig3}
the real and imaginary parts of the corresponding
two-channel-weighted photoionization amplitude $T_{3d}(\omega)$  along with those of the amplitude $T_{3d \to f}(\omega)$.
One can see from Fig.~\ref{fig3} that, near $48$ eV, the relative behavior of ${\rm Re}T_{3d \to f}$  and ${\rm Im}T_{3d \to f}$ of the partial $T_{3d \to f}$ amplitude is quite different than
the relative behavior of ${\rm Re}T_{3d}$  and ${\rm Im}T_{3d}$ of the total amplitude $T_{3d}$. Indeed, ${\rm Re}T_{3d \to f}$ and ${\rm Im}T_{3d \rightarrow f}$ take the zero values simultaneously,
i.e., at the same energy $\omega \approx 48$ eV, so their ratio and, thus, partial $\phi_{3d \rightarrow f}$ and $\tau_{3d \rightarrow f}$ remain about constant through 48 eV as well. In contrast, the
zeros of ${\rm Re}T_{3d}$  and ${\rm Im}T_{3d}$ are separated from each other by
the energy interval of about $1$ eV inside of which ${\rm Re}T_{3d}$ and ${\rm Im}T_{3d}$ take equal values at $\omega \approx 48$ eV. This causes a clear variation, which appears to be resonant,
in $\tan\varphi(\omega)$ and, thus, in the phase $\varphi(\omega)$, and time delay $\tau_{3d}(\omega)$ as well. As a result, $\tau_{3d}(\omega)$ possesses the additional strong $48$-eV resonance but $\tau_{3d \to f}(\omega)$ does not. This finding is both interesting and important since it reveals the necessity for accounting for both a generally dominant ($3d \to f$) and a generally weaker ($3d \to p$) transitions
 in photoionization time delay calculations.

\subsection{Mn $4s$-photoionization}

In this calculation, we use the experimental values \cite{0034-4885-55-7-002}
$I_{3d\uparrow}$($^{5}D_{4})=14.301$~eV (versus $I_{3d\uparrow}^{\rm SPHF}=17.43$~eV),
$I_{4s\uparrow}$($^{5}S)=8.611$~eV (versus $I_{4s\uparrow}^{\rm SPHF}=7.44$~eV), and
$I_{4s\downarrow}$($^{7}S)=7.431$~eV (versus $I_{4s\downarrow}^{\rm SPHF}=6.15$~eV).  This
is because the $4s$$\uparrow$- and $4s$$\downarrow$-subshells of the
Mn atom are significantly closer, in terms of energy, to the
multielectron $3d^{5}$$\uparrow$-subshell than predicted by the SPHF
theory. The use of the calculated SPHF ionization potentials in this
case would have resulted in an underestimated coupling between the
$4s$- and $3d$-ionization channels. Note that, as in the above case
of the $3d$-photoionization, the calculated SPRPAE data for the
$4s$-photoionization amplitudes were obtained by a direct solution of
the SPRPAE equations, in contrast to work \cite{JPB88,JPB90} where a
Fano single-resonance formalism was exploited.

\begin{figure}
\includegraphics[width=8cm]{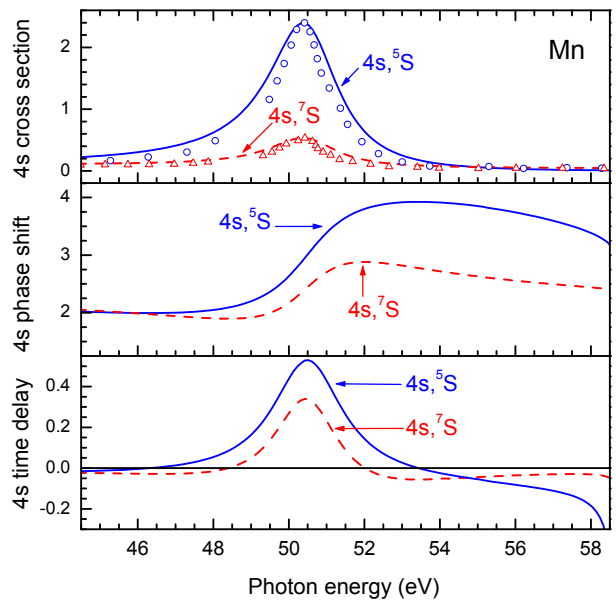}
\caption{(Color online) Calculated SPRPAE (present work) and
experimentally measured \cite{Whitfield94} (open circles and
triangles) Mn $4s$-photoionization cross sections (in units of Mb)
$\sigma_{4s\uparrow}$($^{5}S)$ and $\sigma_{4s\downarrow}$($^{7}S)$,
as well as calculated SPRPAE phase shifts (in units of radians)
$\varphi_{4s\uparrow}$($^{5}S)$ and $\varphi_{4s\downarrow}$($^{7}S)$
of the corresponding photoionization amplitudes,
Eq.~(\ref{amplitude}), and time delays (in units of femtoseconds)
$\tau_{4s\uparrow}$($^{5}S)$, and $\tau_{4s\downarrow}$($^{7}S)$, as
marked.  Relative experimental data \cite{Whitfield94} were normalized
to the maxima in the calculated SPRPAE $\sigma_{4s\uparrow}$($^{5}S)$
and
$\sigma_{4s\downarrow}$($^{7}S)$, respectively, and were shifted by
$0.28$ eV towards higher energies to match the position of the maxima
in
the calculated cross sections.}
\label{fig4}
\end{figure}

Results of the present SPRPAE calculation of
$\sigma_{4s\uparrow}$($^{5}S)$, $\sigma_{4s\downarrow}$($^{7}S)$,
$\varphi_{4s\uparrow}$($^{5}S)$, $\varphi_{4s\downarrow}$($^{7}S)$, as
well as $\tau_{4s\uparrow}$($^{5}S)$ and
$\tau_{4s\downarrow}$($^{7}S)$ are depicted in Fig.~\ref{fig4}.
Note the good agreement between experiment and theory for
$\sigma_{4s\uparrow}$($^{5}S)$ and
$\sigma_{4s\downarrow}$($^{7}S)$. Furthermore, note how the time
delays $\tau_{4s\uparrow}$($^{5}S)$ and $\tau_{4s\downarrow}$($^{7}S)$
are dramatically increased in the resonance region. Next, note how
$\tau_{4s\uparrow}$($^{5}S)$ and $\tau_{4s\downarrow}$($^{7}S)$ differ
strongly from each other in this energy region where $\tau_{4s\uparrow}$($^{5}S)$
exceeds $\tau_{4s\downarrow}$($^{7}S)$ everywhere below an energy of $\omega \approx 54.7$ eV;
above this energy the situation changes to the opposite. It is also interesting to note that
 $\tau_{4s\uparrow}$($^{5}S)$ and $\tau_{4s\downarrow}$($^{7}S)$ differ from each other strongly
not only by magnitude but by sign as well, depending on $\omega$.
The noted differences between $\tau_{4s\uparrow}$($^{5}S)$ and $\tau_{4s\downarrow}$($^{7}S)$ constitute
an interesting finding of a strong term-dependence of time delay
in $4s$-photoionization of Mn. This is due to SPRPAE exchange diagrams (d) and (f) of Fig.~\ref{fig1} which
contribute to photoionization of $4s$$\uparrow$-electron differently than to the $4s$$\downarrow$-electron.
For
instance (see Fig.~\ref{fig1}), if the intermediate excitation ``$j-k'$''$ \equiv 3d$$\uparrow$ $\to \epsilon l$$\uparrow$,
then the exchange diagrams (d) and (f)  affect photoionization of a $4s$$\uparrow$-electron, but not a $4s$$\downarrow$-electron (otherwise the Coulomb interaction would have caused
a spin-flip transition).
Alternatively, the diagrams (d) and (f) of Fig.~\ref{fig1} do not affect the $4s$$\uparrow$-electron but do affect the $4s$$\downarrow$-photoionization amplitude
 when ``$j-k'$''$ \equiv 3p$$\downarrow$~$\to 3d$$\downarrow$. Thus, in the region of the $3p$$\downarrow$ $\to$ $3d$$\downarrow$ giant resonance, where the impact of interchannel coupling
 of the $3p$$\downarrow$ $\to$ $3d$$\downarrow$ and $3d$$\uparrow$ $\to$ $p$$\uparrow$, $f$$\uparrow$ channels with each of the $4s$$\uparrow$ $\to$ $p$$\uparrow$ and $4s$$\downarrow$ $\to$ $p$$\downarrow$
 channels is significant, time delay in the $4s$-photoionization of Mn becomes strongly term dependent.
This should be a general feature of other transition-metal atoms and ions as well.

\section{Conclusion}

It has been demonstrated in this paper that time delays of the
corresponding Mn $3d$- and $4s$-photoionization channels are
dramatically increased in the region of the $3p \to 3d$ giant
autoionization resonance.
Furthermore, by utilizing the Fano formalism, it has been shown
that the photoionization time delay in the autoionization resonance region
is explicitly associated with the resonance lifetime. The latter, thus, can be determined directly
from results of attosecond time delay experiments.
Furthermore, it has been found that the time
delay of the $4s$-photoionization channel is strongly term-dependent,
resulting in significant differences between time delays
$\tau_{4s\uparrow}$($^{5}S)$ and $\tau_{4s\downarrow}$($^{7}S)$.
Strong maxima in $\tau_{3d\uparrow}$($^{4}D)$, $\tau_{4s\uparrow}$($^{5}S)$, and
$\tau_{4s\downarrow}$($^{7}S)$ emerge at
photon energies where the corresponding photoionization cross-sections
are large, particularly for $3d$-photoionization.
This should simplify
experimental measurements of the phenomena described in Mn.
Moreover,
the $3p \to 3d$ giant autoionization resonance is known to occur
in the photoabsorption spectra of Mn$^{+}$, metallic Mn, molecular MnCl$_{2}$, and
solid MnCl$_{2}$ as well \cite{Sonntag1969597,Costello91}. This
provides the flexibility for experimental
verification of the predictions made in this paper. It is
expected that the features unraveled in time delays of the Mn
photoelectron emission channels will emerge in other $3d$- and
$4d$-transitions elements  and rare-earths where giant
autoionization resonances exist as well. In other words, the
 features of time delays in Mn
photoionization, unveiled in the present paper, are, in fact, inherent
properties of not only the Mn
atom, in particular, but other transition-metal and rare-earth atoms,  in
general.  Correspondingly, results of the present paper provide
guidance into photoionization time delays in those atoms as well.

Next, it is important to note that the well-known giant resonance in
$4d$-photoionization of Xe, see, e.g., Ref.~\cite{AC97}, has found an
important application for the induction of a strong enhancement of the
high harmonic generation (HHG) process \cite{Shiner2011}. The autoionization
multielectron dynamics can be probed by the HHG
technique \cite{PhysRevLett.104.123901} as well, particularly in the $3p \to 3d$ giant resonance
in Mn,  as in recent work by Ganeev et.~al. \cite{Ganeev}, in view of the large value of the Mn
$3d$-photoionization cross section in there; the latter is even greater than the
Xe $4d$- photoionization cross section. Although experiments with
metal vapors and ablation plasmas may be more challenging than with
noble gases, HHG experiments can now be performed \cite{Ghimire2011}
and analyzed \cite{PhysRevLett.113.073901} in the condensed matter phase
as well. Solid state Mn is a promising target for such a study.

In conclusion, it would be interesting to study how
accounting for a goodly number of other open channels in the Mn $3p \to 3d$ giant resonance region,
omitted in our SPRPAE study, could modify the predicted gross features in $3d$- and $4s$-time delays
in this atom (apparently, it does not significantly modify the
 calculated SPRPAE $\sigma_{3d}$ or
$\beta_{3d}$ in Mn). The authors' expectation is that possible alterations should not be drastic, but in the
absence of experiment or other theoretical calculations on this subject the question basically remains open.
In fact, we consider the photoionization time delay phenomenon in highly correlated atoms as a novel
touchstone for a finer testing of existing and to-be-developed many-body theories against experiment.
We urge the development of such calculations and experiments, and we hope that the provided in the present paper
initial insight into
time delay in the Mn photoionization  serves as the impetus for such development.

\begin{acknowledgments}
The authors are grateful to Dr. Vladislav Yakovlev for his interest in this study.
Dr. Jobin Jose is thanked for his assistance with the calculations.
V.K.D.~acknowledges  the support of NSF under grant No.~PHY-$1305085$.
S.T.M.~acknowledges the support of the Chemical Sciences,
Geosciences and Biosciences Division, Office of Basic Energy
Sciences, Office of Science, US Department of Energy under
Grant No.~DE-FG02-03ER15428.
P.C.D.~appreciates the support of the grant from the Department of Science and
Technology, Government of India.
\end{acknowledgments}


\begin{thebibliography}{33}
\expandafter\ifx\csname natexlab\endcsname\relax\def\natexlab#1{#1}\fi
\expandafter\ifx\csname bibnamefont\endcsname\relax
  \def\bibnamefont#1{#1}\fi
\expandafter\ifx\csname bibfnamefont\endcsname\relax
  \def\bibfnamefont#1{#1}\fi
\expandafter\ifx\csname citenamefont\endcsname\relax
  \def\citenamefont#1{#1}\fi
\expandafter\ifx\csname url\endcsname\relax
  \def\url#1{\texttt{#1}}\fi
\expandafter\ifx\csname urlprefix\endcsname\relax\def\urlprefix{URL }\fi
\providecommand{\bibinfo}[2]{#2}
\providecommand{\eprint}[2][]{\url{#2}}

\bibitem{M.Schultze06252010} M. Schultze,
M. Fie\ss, N. Karpowicz,
J. Gagnon, M. Korbman, M. Hofstetter, S. Neppl, A. L. Cavalieri,
Y. Komninos, Th. Mercouris, C. A. Nicolaides, R. Pazourek, S. Nagele,
J. Feist, J. Burgd\"{o}rfer, A. M. Azzeer, R. Ernstorfer,
R. Kienberger, U. Kleineberg, E. Goulielmakis, F. Krausz, and
V. S. Yakovlev,
%
\textit{Delay in Photoemission}, Science \textbf{328}, 1658 (2010).

\bibitem{PhysRevLett.106.143002} K. Kl\"{u}nder,
J. M. Dahlstr\"{o}m,
M. Gisselbrecht, T. Fordell, M. Swoboda, D. Gu\'enot, P. Johnsson,
J. Caillat, J. Mauritsson, A. Maquet, R. Ta\"{\i}eb, and
A. L'Huillier,
%
\textit{Probing single-photon ionization on the
attosecond time scale}, Phys. Rev. Lett. \textbf{106}, 143002 (2011).

\bibitem{PhysRevA.85.053424} D. Gu\'enot, K. Kl\"{u}nder,
C. L. Arnold, D. Kroon, J. M. Dahlstr\"{o}m, M. Miranda, T. Fordell,
M. Gisselbrecht, P. Johnsson, J. Mauritsson, E. Lindroth, A. Maquet,
R. Ta\"{\i}eb, A. L'Huillier, and A. S. Kheifets,
\textit{Photoemission-time-delay measurements and calculations close
to the $3s$-ionization-cross-section minimum in Ar}, Phys. Rev. A
\textbf{85}, 053424 (2012).

\bibitem{Guenot2014} D. Gu\'enot, D. Kroon, E. Balogh, E. W. Larsen, M. Kotur,
M. Miranda, T. Fordell, P. Johnsson, J. Mauritsson, M. Gisselbrecht, K. Varj\'u,
C. L. Arnold, T. Carette, A. S. Kheifets, E. Lindroth, A. L'Huillier, and J. M. Dahlstr\"{o}m,
\textit{Measurements of photoemission time delays in noble gas atoms}, J. Phys. B \textbf{47}, $245602$ (2014).

\bibitem{DiMauro2014} Caryn Palatchi, J. M. Dahlstr\"{o}m, A. S. Kheifets, I. A. Ivanov, D. M. Canaday,
P. Agostini, and L. F. DiMauro, \textit{Atomic delay in helium, neon, argon and krypton},
J. Phys. B \textbf{47}, 245003 (2014).

\bibitem[{\citenamefont{Schoun et~al.}(2014)\citenamefont{Schoun, Chirla,
  Wheeler, Roedig, Agostini, DiMauro, Schafer, and
  Gaarde}}]{PhysRevLett.112.153001}
\bibinfo{author}{\bibfnamefont{S.~B.} \bibnamefont{Schoun}},
  \bibinfo{author}{\bibfnamefont{R.}~\bibnamefont{Chirla}},
  \bibinfo{author}{\bibfnamefont{J.}~\bibnamefont{Wheeler}},
  \bibinfo{author}{\bibfnamefont{C.}~\bibnamefont{Roedig}},
  \bibinfo{author}{\bibfnamefont{P.}~\bibnamefont{Agostini}},
  \bibinfo{author}{\bibfnamefont{L.~F.} \bibnamefont{DiMauro}},
  \bibinfo{author}{\bibfnamefont{K.~J.} \bibnamefont{Schafer}},
  \bibnamefont{and} \bibinfo{author}{\bibfnamefont{M.~B.}
  \bibnamefont{Gaarde}}, \emph{\bibinfo{title}{Attosecond pulse shaping around
  a {Cooper} minimum}}, \bibinfo{journal}{Phys. Rev. Lett.}
  \textbf{\bibinfo{volume}{112}}, \bibinfo{pages}{153001}
  (\bibinfo{year}{2014}).

\bibitem[{\citenamefont{Kheifets and Ivanov}(2010)}]{PhysRevLett.105.233002}
\bibinfo{author}{\bibfnamefont{A.~S.} \bibnamefont{Kheifets}} \bibnamefont{and}
  \bibinfo{author}{\bibfnamefont{I.~A.} \bibnamefont{Ivanov}},
  \emph{\bibinfo{title}{Delay in atomic photoionization}},
  \bibinfo{journal}{Phys. Rev. Lett.}
  \textbf{\bibinfo{volume}{105}}, \bibinfo{pages}{233002}
  (\bibinfo{year}{2010}).

\bibitem[{\citenamefont{Moore et~al.}(2011)\citenamefont{Moore, Lysaght,
  Parker, van~der Hart, and Taylor}}]{PhysRevA.84.061404}
\bibinfo{author}{\bibfnamefont{L.~R.} \bibnamefont{Moore}},
  \bibinfo{author}{\bibfnamefont{M.~A.} \bibnamefont{Lysaght}},
  \bibinfo{author}{\bibfnamefont{J.~S.} \bibnamefont{Parker}},
  \bibinfo{author}{\bibfnamefont{H.~W.} \bibnamefont{van~der Hart}},
  \bibnamefont{and} \bibinfo{author}{\bibfnamefont{K.~T.}
  \bibnamefont{Taylor}}, \emph{\bibinfo{title}{Time delay between photoemission
  from the $2p$ and $2s$ subshells of neon}}, \bibinfo{journal}{Phys. Rev. A}
  \textbf{\bibinfo{volume}{84}}, \bibinfo{pages}{061404}
  (\bibinfo{year}{2011}).

\bibitem[{\citenamefont{Nagele et~al.}(2011)\citenamefont{Nagele, Pazourek,
  Feist, Doblhoff-Dier, Lemell, T\"ok\'esi, and
  Burgd\"orfer}}]{0953-4075-44-8-081001}
\bibinfo{author}{\bibfnamefont{S.}~\bibnamefont{Nagele}},
  \bibinfo{author}{\bibfnamefont{R.}~\bibnamefont{Pazourek}},
  \bibinfo{author}{\bibfnamefont{J.}~\bibnamefont{Feist}},
  \bibinfo{author}{\bibfnamefont{K.}~\bibnamefont{Doblhoff-Dier}},
  \bibinfo{author}{\bibfnamefont{C.}~\bibnamefont{Lemell}},
  \bibinfo{author}{\bibfnamefont{K.}~\bibnamefont{T\"ok\'esi}},
  \bibnamefont{and}
  \bibinfo{author}{\bibfnamefont{J.}~\bibnamefont{Burgd\"orfer}},
  \emph{\bibinfo{title}{Time-resolved photoemission by attosecond streaking:
  extraction of time information}}, \bibinfo{journal}{J.~Phys.~B}
  \textbf{\bibinfo{volume}{44}}(\bibinfo{number}{8}), \bibinfo{pages}{081001}
  (\bibinfo{year}{2011}).


\bibitem[{\citenamefont{Dahlstr\"om et~al.}(2012)\citenamefont{Dahlstr\"om,
  Carette, and Lindroth}}]{PhysRevA.86.061402}
\bibinfo{author}{\bibfnamefont{J.~M.} \bibnamefont{Dahlstr\"om}},
  \bibinfo{author}{\bibfnamefont{T.}~\bibnamefont{Carette}}, \bibnamefont{and}
  \bibinfo{author}{\bibfnamefont{E.}~\bibnamefont{Lindroth}},
  \emph{\bibinfo{title}{Diagrammatic approach to attosecond delays in
  photoionization}}, \bibinfo{journal}{Phys. Rev. A}
  \textbf{\bibinfo{volume}{86}}, \bibinfo{pages}{061402}
  (\bibinfo{year}{2012}).


\bibitem{Dahlstrom2012} J. M. Dahlstr\"{o}m, A. L'Huillier, and
A. Maquet, \textit{Introduction to attosecond delays in
photoionization}, J. Phys. B \textbf{45}, 183001 (2012).

\bibitem{Dahlstroem2012b} J. Dahlstr\"{o}m, D. Gu\'enot, K. Kl\"{u}nder, M. Gisselbrecht,
J. Mauritsson, A. L. Huillier, A. Maquet, and R. Ta\"{\i}ieb, \textit{Theory of attosecond delays in laser-assisted photoionization},
Chem. Phys. \textbf{414}, 53 (2012).

\bibitem[{\citenamefont{Kheifets}(2013)}]{PhysRevA.87.063404}
\bibinfo{author}{\bibfnamefont{A.~S.} \bibnamefont{Kheifets}},
  \emph{\bibinfo{title}{Time delay in valence-shell photoionization of
  noble-gas atoms}}, \bibinfo{journal}{Phys. Rev. A}
  \textbf{\bibinfo{volume}{87}}, \bibinfo{pages}{063404}
  (\bibinfo{year}{2013}).

\bibitem[{\citenamefont{Feist et~al.}(2014)\citenamefont{Feist, Zatsarinny,
  Nagele, Pazourek, Burgd\"orfer, Guan, Bartschat, and
  Schneider}}]{PhysRevA.89.033417}
\bibinfo{author}{\bibfnamefont{J.}~\bibnamefont{Feist}},
  \bibinfo{author}{\bibfnamefont{O.}~\bibnamefont{Zatsarinny}},
  \bibinfo{author}{\bibfnamefont{S.}~\bibnamefont{Nagele}},
  \bibinfo{author}{\bibfnamefont{R.}~\bibnamefont{Pazourek}},
  \bibinfo{author}{\bibfnamefont{J.}~\bibnamefont{Burgd\"orfer}},
  \bibinfo{author}{\bibfnamefont{X.}~\bibnamefont{Guan}},
  \bibinfo{author}{\bibfnamefont{K.}~\bibnamefont{Bartschat}},
  \bibnamefont{and} \bibinfo{author}{\bibfnamefont{B.~I.}
  \bibnamefont{Schneider}}, \emph{\bibinfo{title}{Time delays for attosecond
  streaking in photoionization of neon}}, \bibinfo{journal}{Phys. Rev. A}
  \textbf{\bibinfo{volume}{89}}, \bibinfo{pages}{033417}
  (\bibinfo{year}{2014}).

\bibitem[{\citenamefont{Saha et~al.}(2014)\citenamefont{Saha, Mandal, Jose,
  Varma, Deshmukh, Kheifets, Dolmatov, and Manson}}]{PhysRevA.90.053406}
\bibinfo{author}{\bibfnamefont{S.}~\bibnamefont{Saha}},
  \bibinfo{author}{\bibfnamefont{A.}~\bibnamefont{Mandal}},
  \bibinfo{author}{\bibfnamefont{J.}~\bibnamefont{Jose}},
  \bibinfo{author}{\bibfnamefont{H.~R.} \bibnamefont{Varma}},
  \bibinfo{author}{\bibfnamefont{P.~C.} \bibnamefont{Deshmukh}},
  \bibinfo{author}{\bibfnamefont{A.~S.} \bibnamefont{Kheifets}},
  \bibinfo{author}{\bibfnamefont{V.~K.} \bibnamefont{Dolmatov}},
  \bibnamefont{and} \bibinfo{author}{\bibfnamefont{S.~T.}
  \bibnamefont{Manson}}, \emph{\bibinfo{title}{Relativistic effects in
  photoionization time delay near the {Cooper} minimum of noble-gas atoms}},
  \bibinfo{journal}{Phys. Rev. A} \textbf{\bibinfo{volume}{90}},
  \bibinfo{pages}{053406} (\bibinfo{year}{2014}).

\bibitem[{\citenamefont{Eisenbud}(1948)}]{Eisenbud1948}
\bibinfo{author}{\bibfnamefont{L.}~\bibnamefont{Eisenbud}}, \textit{Formal Properties of Nuclear Collisions}, Ph.D. thesis,
  \bibinfo{school}{Princeton University} (\bibinfo{year}{1948}).

\bibitem[{\citenamefont{Wigner}(1955)}]{PhysRev.98.145}
\bibinfo{author}{\bibfnamefont{E.~P.} \bibnamefont{Wigner}},
  \emph{\bibinfo{title}{Lower limit for the energy derivative of the scattering
  phase shift}}, \bibinfo{journal}{Phys. Rev.}
  \textbf{\bibinfo{volume}{98}}, \bibinfo{pages}{145}
  (\bibinfo{year}{1955}).

\bibitem[{\citenamefont{K. et~al.}(2013)\citenamefont{Dolmatov, Amusia, and Chernysheva
  }}]{PhysRevA.88.042706}
\bibinfo{author}{\bibfnamefont{V.~K.} \bibnamefont{Dolmatov}},
  \bibinfo{author}{\bibfnamefont{M.~Ya.} \bibnamefont{Amusia}}, \bibnamefont{and}
  \bibinfo{author}{\bibfnamefont{L.~V.} \bibnamefont{Chernysheva}},
  \emph{\bibinfo{title}{Electron elastic scattering off a semifilled-shell
  atom: {The Mn} atom}}, \bibinfo{journal}{Phys. Rev. A}
  \textbf{\bibinfo{volume}{88}}, \bibinfo{pages}{042706}
  (\bibinfo{year}{2013}).

\bibitem[{\citenamefont{Connerade et.~al.}(1976)\citenamefont{Connerade,
  Mansfield, and Martin}}]{Connerade1976}
\bibinfo{author}{\bibfnamefont{J.~P.} \bibnamefont{Connerade}},
  \bibinfo{author}{\bibfnamefont{M.~W.~D.} \bibnamefont{Mansfield}},
  \bibnamefont{and} \bibinfo{author}{\bibfnamefont{M.~A.~P.}
  \bibnamefont{Martin}}, \emph{\bibinfo{title}{Observation of a giant resonance
  in the 3p absorption spectrum of {Mn I}}}, \bibinfo{journal}{Proc. R. Soc. London, Ser. A}
  \textbf{\bibinfo{volume}{350}},
  \bibinfo{pages}{405} (\bibinfo{year}{1976}).

\bibitem[{\citenamefont{Sonntag and Zimmermann}(1992)}]{0034-4885-55-7-002}
\bibinfo{author}{\bibfnamefont{B.}~\bibnamefont{Sonntag}} \bibnamefont{and}
  \bibinfo{author}{\bibfnamefont{P.}~\bibnamefont{Zimmermann}},
  \emph{\bibinfo{title}{{XUV} spectroscopy of metal atoms}},
  \bibinfo{journal}{Rep. Prog. Phys.}
  \textbf{\bibinfo{volume}{55}}, \bibinfo{pages}{911}
  (\bibinfo{year}{1992}).

\bibitem[{\citenamefont{Martins et~al.}(2006)\citenamefont{Martins, Godehusen,
  Richter, Wernet, and Zimmermann}}]{0953-4075-39-5-R01}
\bibinfo{author}{\bibfnamefont{M.}~\bibnamefont{Martins}},
  \bibinfo{author}{\bibfnamefont{K.}~\bibnamefont{Godehusen}},
  \bibinfo{author}{\bibfnamefont{T.}~\bibnamefont{Richter}},
  \bibinfo{author}{\bibfnamefont{P.}~\bibnamefont{Wernet}}, \bibnamefont{and}
  \bibinfo{author}{\bibfnamefont{P.}~\bibnamefont{Zimmermann}},
  \emph{\bibinfo{title}{Open shells and multi-electron interactions: core level
  photoionization of the 3d metal atoms}}, \bibinfo{journal}{J. Phys. B}
  \textbf{\bibinfo{volume}{39}}, \bibinfo{pages}{R79}
  (\bibinfo{year}{2006}).
%
\bibitem{Athens} Y. Komninos, Th. Mercouris, and C. A. Nicolaides,
\textit{Regular series of doubly excited states inside two-electron
continua: Application to $2s^{2}$-hole states in Neon, above the
Ne$^{2+}$ $1s^{2}2s^{2}2p^{4}$ and $1s^{2}2s2p^{5}$ thresholds},
Phys. Rev. A \textbf{83}, 022501 (2011).
%
\bibitem[{\citenamefont{Carette et~al.}(2013)\citenamefont{Carette,
  Dahlstr\"om, Argenti, and Lindroth}}]{PhysRevA.87.023420}
\bibinfo{author}{\bibfnamefont{T.}~\bibnamefont{Carette}},
  \bibinfo{author}{\bibfnamefont{J.~M.} \bibnamefont{Dahlstr\"om}},
  \bibinfo{author}{\bibfnamefont{L.}~\bibnamefont{Argenti}},
  \bibnamefont{and}
  \bibinfo{author}{\bibfnamefont{E.}~\bibnamefont{Lindroth}},
  \emph{\bibinfo{title}{Multiconfigurational Hartree-Fock
  close-coupling
  ansatz: Application to the argon photoionization cross section and
  delays}},
  \bibinfo{journal}{Phys. Rev. A}
\href{http://link.aps.org/doi/10.1103/PhysRevA.87.023420}{
\textbf{\bibinfo{volume}{87}},
  \bibinfo{pages}{023420} (\bibinfo{year}{2013})}.
%
\bibitem[{\citenamefont{{Sabbar} et~al.}(2014)\citenamefont{{Sabbar}, {Heuser},
  {Boge}, {Lucchini}, {Carette}, {Lindroth}, {Gallmann}, {Cirelli}, and
  {Keller}}}]{2014arXiv1407.6623S}
\bibinfo{author}{\bibfnamefont{M.}~\bibnamefont{{Sabbar}}},
  \bibinfo{author}{\bibfnamefont{S.}~\bibnamefont{{Heuser}}},
  \bibinfo{author}{\bibfnamefont{R.}~\bibnamefont{{Boge}}},
  \bibinfo{author}{\bibfnamefont{M.}~\bibnamefont{{Lucchini}}},
  \bibinfo{author}{\bibfnamefont{T.}~\bibnamefont{{Carette}}},
  \bibinfo{author}{\bibfnamefont{E.}~\bibnamefont{{Lindroth}}},
  \bibinfo{author}{\bibfnamefont{L.}~\bibnamefont{{Gallmann}}},
  \bibinfo{author}{\bibfnamefont{C.}~\bibnamefont{{Cirelli}}},
  \bibnamefont{and} \bibinfo{author}{\bibfnamefont{U.}~\bibnamefont{{Keller}}},
  \emph{\bibinfo{title}{{Resonance effects in photoemission time delays}}},
  \bibinfo{journal}{ArXiv e-prints}  (\bibinfo{year}{2014}),
  \eprint{1407.6623}.
%

\bibitem[{\citenamefont{Krause et~al.}(1984)\citenamefont{Krause, Carlson, and
  Fahlman}}]{PhysRevA.30.1316}
\bibinfo{author}{\bibfnamefont{M.~O.} \bibnamefont{Krause}},
  \bibinfo{author}{\bibfnamefont{T.~A.} \bibnamefont{Carlson}},
  \bibnamefont{and} \bibinfo{author}{\bibfnamefont{A.}~\bibnamefont{Fahlman}},
  \emph{\bibinfo{title}{Photoelectron spectrometry of manganese vapor between
  12 and 110 {eV}}}, \bibinfo{journal}{Phys. Rev. A}
  \textbf{\bibinfo{volume}{30}}, \bibinfo{pages}{1316} (\bibinfo{year}{1984}).

\bibitem{Whitfield94} S. B. Whitfield, M. O. Krause, P. van der Meulen, and C. D. Caldwell,
\textit{High-resolution photoelectron spectrometry of atomic manganese from the region of
the $3p \to 3d$ giant resonance to $120$ eV},
Phys. Rev. A \textbf{50}, 1269 (1994).

\bibitem[{\citenamefont{Osawa et~al.}(2012)\citenamefont{Osawa, Kawajiri,
  Suzuki, Nagata, Azuma, and Koike}}]{0953-4075-45-22-225204}
\bibinfo{author}{\bibfnamefont{T.}~\bibnamefont{Osawa}},
  \bibinfo{author}{\bibfnamefont{K.}~\bibnamefont{Kawajiri}},
  \bibinfo{author}{\bibfnamefont{N.}~\bibnamefont{Suzuki}},
  \bibinfo{author}{\bibfnamefont{T.}~\bibnamefont{Nagata}},
  \bibinfo{author}{\bibfnamefont{Y.}~\bibnamefont{Azuma}}, \bibnamefont{and}
  \bibinfo{author}{\bibfnamefont{F.}~\bibnamefont{Koike}},
  \emph{\bibinfo{title}{Photoion-yield study of the $3p-3d$ giant resonance
  excitation region of isolated {Cr}, {Mn} and {Fe} atoms}},
  \bibinfo{journal}{J. Phys. B}
  \textbf{\bibinfo{volume}{45}}, \bibinfo{pages}{225204}
  (\bibinfo{year}{2012}).


\bibitem[{\citenamefont{Amus'ya et~al.}(1983)\citenamefont{Amus'ya, Dolmatov,
  and lvanov}}]{JETP1983}
\bibinfo{author}{\bibfnamefont{M.~Ya.} \bibnamefont{Amus'ya}},
  \bibinfo{author}{\bibfnamefont{V.~K.}~\bibnamefont{Dolmatov}}, \bibnamefont{and}
  \bibinfo{author}{\bibfnamefont{V.~K.}~\bibnamefont{lvanov}},
  \emph{\bibinfo{title}{Photoionization of atoms with half-filled shells}},
  \bibinfo{journal}{Sov. Phys. -- JETP}
  \textbf{\bibinfo{volume}{58}}, \bibinfo{pages}{67}
  (\bibinfo{year}{1983}).


\bibitem[{\citenamefont{Amusia and Dolmatov}(1993)}]{0953-4075-26-8-010}
\bibinfo{author}{\bibfnamefont{M.~Ya.} \bibnamefont{Amusia}} \bibnamefont{and}
  \bibinfo{author}{\bibfnamefont{V.~K.} \bibnamefont{Dolmatov}},
  \emph{\bibinfo{title}{Photoionization of inner ns electrons in semifilled
  shell atoms ($3s$ electrons in a {Mn} atom)}}, \bibinfo{journal}{J. Phys. B}
  \textbf{\bibinfo{volume}{26}}, \bibinfo{pages}{1425}
  (\bibinfo{year}{1993}).

\bibitem[{\citenamefont{Amusia and Chernysheva}(1997)}]{AC97}
\bibinfo{author}{\bibfnamefont{M.~Ya.} \bibnamefont{Amusia}} \bibnamefont{and}
  \bibinfo{author}{\bibfnamefont{L.~V.} \bibnamefont{Chernysheva}},
  \emph{\bibinfo{title}{Computation of atomic processes : {A} Handbook for the
  {ATOM} Programs}} (\bibinfo{publisher}{IOP},
  \bibinfo{address}{Bristol}, \bibinfo{year}{1997}).

\bibitem{JPB88} M. Ya. Amusia, V. K. Dolmatov, and V. M. Romanenko, \textit{Exchange electron correlation effects in outer-level
photoionisation of half-filled subshell atoms (Mn)},
J. Phys. B. \textbf{21}, L151 (1988).

\bibitem{JPB90} M. Ya. Amusia, V. K. Dolmatov, and M. M. Mansurov, \textit{A new feature of the $3p \to 3d$ transition in the Mn atom},
J. Phys. B \textbf{23}, L491 (1990).

\bibitem[{\citenamefont{Garvin et~al.}(1983)\citenamefont{Garvin, Brown,
  Carter, and Kelly}}]{0022-3700-16-9-004}
\bibinfo{author}{\bibfnamefont{L.~J.} \bibnamefont{Garvin}},
  \bibinfo{author}{\bibfnamefont{E.~R.} \bibnamefont{Brown}},
  \bibinfo{author}{\bibfnamefont{S.~L.} \bibnamefont{Carter}},
  \bibnamefont{and} \bibinfo{author}{\bibfnamefont{H.~P.} \bibnamefont{Kelly}},
  \emph{\bibinfo{title}{Calculation of photoionisation cross sections,
  resonance structure and angular distribution for Mn I by many-body
  perturbation theory}}, \bibinfo{journal}{J. Phys. B}
  \textbf{\bibinfo{volume}{16}}, \bibinfo{pages}{L269}
  (\bibinfo{year}{1983}).


\bibitem[{\citenamefont{Slater}(1974)}]{slater1974self}
\bibinfo{author}{\bibfnamefont{J.~C.}~\bibnamefont{Slater}},
  \emph{\bibinfo{title}{The Self-Consistent Field for Molecules and Solids}}
 (\bibinfo{publisher}{McGraw-Hill, New York}, \bibinfo{year}{1974}).


\bibitem{DolmJPB96} V. K. Dolmatov and M. M. Mansurov, \textit{Trends in valence autoionizing Rydbergs for $nd^{5}$
semifilled-shell ground and excited atoms and ions: Cr,
Mn, Mn$^{+}$, Fe$^{+}$, Fe$^{2+}$, Mo, Tc, Tc$^{+}$ and Re}, J. Phys. B \textbf{29}, L307 (1996).

\bibitem{DolmPRA06} V. K. Dolmatov and S. T. Manson,
\textit{Strong final-state term dependence of nondipole photoelectron angular distributions
from half-filled shell atoms}, Phys. Rev. A \textbf{74}, 032705 (2006).


\bibitem[{\citenamefont{Smith}(1960)}]{PhysRev.118.349}
\bibinfo{author}{\bibfnamefont{F.~T.} \bibnamefont{Smith}},
  \emph{\bibinfo{title}{Lifetime matrix in collision theory}},
  \bibinfo{journal}{Phys. Rev.} \textbf{\bibinfo{volume}{118}},
  \bibinfo{pages}{349} (\bibinfo{year}{1960}).

\bibitem[{\citenamefont{W\"atzel et~al.}(2015)\citenamefont{W\"atzel,
  Moskalenko, Pavlyukh, and Berakdar}}]{0953-4075-48-2-025602}
\bibinfo{author}{\bibfnamefont{J.}~\bibnamefont{W\"atzel}},
  \bibinfo{author}{\bibfnamefont{A.~S.} \bibnamefont{Moskalenko}},
  \bibinfo{author}{\bibfnamefont{Y.}~\bibnamefont{Pavlyukh}}, \bibnamefont{and}
  \bibinfo{author}{\bibfnamefont{J.}~\bibnamefont{Berakdar}},
  \emph{\bibinfo{title}{Angular resolved time delay in photoemission}},
  \bibinfo{journal}{J. Phys. B} \textbf{\bibinfo{volume}{48}}(\bibinfo{number}{2}),
  \bibinfo{pages}{025602} (\bibinfo{year}{2015}).

\bibitem[{\citenamefont{Dahlstr\"om and Lindroth}(2014)}]{Dahlstrom2014}
\bibinfo{author}{\bibfnamefont{J.~M.} \bibnamefont{Dahlstr\"om}}
  \bibnamefont{and} \bibinfo{author}{\bibfnamefont{E.}~\bibnamefont{Lindroth}},
  \emph{\bibinfo{title}{Study of attosecond delays using perturbation diagrams
  and exterior complex scaling}}, \bibinfo{journal}{J. Phys. B}
  \textbf{\bibinfo{volume}{47}}, \bibinfo{pages}{124012}
  (\bibinfo{year}{2014}).

\bibitem[{\citenamefont{Jim\'enez-Gal\'an
  et~al.}(2014)\citenamefont{Jim\'enez-Gal\'an, Argenti, and
  Mart\'{i}n}}]{PhysRevLett.113.263001}
\bibinfo{author}{\bibfnamefont{A.}~\bibnamefont{Jim\'enez-Gal\'an}},
  \bibinfo{author}{\bibfnamefont{L.}~\bibnamefont{Argenti}}, \bibnamefont{and}
  \bibinfo{author}{\bibfnamefont{F.}~\bibnamefont{Mart\'{i}n}},
  \emph{\bibinfo{title}{Modulation of attosecond beating in resonant two-photon
  ionization}}, \bibinfo{journal}{Phys. Rev. Lett.}
  \textbf{\bibinfo{volume}{113}}, \bibinfo{pages}{263001}
  (\bibinfo{year}{2014}).

\bibitem[{\citenamefont{Fano}(1961)}]{PhysRev.124.1866}
\bibinfo{author}{\bibfnamefont{U.}~\bibnamefont{Fano}},
  \emph{\bibinfo{title}{Effects of configuration interaction on intensities and
  phase shifts}}, \bibinfo{journal}{Phys. Rev.} \textbf{\bibinfo{volume}{124}},
  \bibinfo{pages}{1866} (\bibinfo{year}{1961}).



\bibitem[{\citenamefont{Sonntag et~al.}(1969)\citenamefont{Sonntag, Haensel,
  and Kunz}}]{Sonntag1969597}
\bibinfo{author}{\bibfnamefont{B.}~\bibnamefont{Sonntag}},
  \bibinfo{author}{\bibfnamefont{R.}~\bibnamefont{Haensel}}, \bibnamefont{and}
  \bibinfo{author}{\bibfnamefont{C.}~\bibnamefont{Kunz}},
  \emph{\bibinfo{title}{Optical absorption measurements of the transition
  metals {Ti}, {V}, {Cr}, {Mn}, {Fe}, {Co}, {Ni} in the region of 3p electron
  transitions}}, \bibinfo{journal}{Solid State Communications}
  \textbf{\bibinfo{volume}{7}}, \bibinfo{pages}{597 }
  (\bibinfo{year}{1969}).

\bibitem{Costello91} J. T. Costello, E. T. Kennedy, B. F. Sonntag, and
C. W. Clark,
\emph{\bibinfo{title}{$3p$ photoabsorption of free and bound Cr, Cr$^{+}$, Mn,
and Mn$^{+}$}}, Phys. Rev. A \textbf{43}, 1441 (1991).

\bibitem[{\citenamefont{Drescher et~al.}(2002)\citenamefont{Drescher,
  Hentschel, Kienberger, Uiberacker, Yakovlev, Scrinzi, Westerwalbesloh,
  Kleineberg, Heinzmann, and Krausz}}]{Drescher2002}
\bibinfo{author}{\bibfnamefont{M.}~\bibnamefont{Drescher}},
  \bibinfo{author}{\bibfnamefont{M.}~\bibnamefont{Hentschel}},
  \bibinfo{author}{\bibfnamefont{R.}~\bibnamefont{Kienberger}},
  \bibinfo{author}{\bibfnamefont{M.}~\bibnamefont{Uiberacker}},
  \bibinfo{author}{\bibfnamefont{V.}~\bibnamefont{Yakovlev}},
  \bibinfo{author}{\bibfnamefont{A.}~\bibnamefont{Scrinzi}},
  \bibinfo{author}{\bibfnamefont{T.}~\bibnamefont{Westerwalbesloh}},
  \bibinfo{author}{\bibfnamefont{U.}~\bibnamefont{Kleineberg}},
  \bibinfo{author}{\bibfnamefont{U.}~\bibnamefont{Heinzmann}},
  \bibnamefont{and} \bibinfo{author}{\bibfnamefont{F.}~\bibnamefont{Krausz}},
  \emph{\bibinfo{title}{Time-resolved atomic inner-shell spectroscopy}},
  \bibinfo{journal}{Nature} \textbf{\bibinfo{volume}{419}},
  \bibinfo{pages}{803} (\bibinfo{year}{2002}).


\bibitem[{\citenamefont{Shiner et~al.}(2011)\citenamefont{Shiner, Schmidt,
  Trallero-Herrero, Worner, Patchkovskii, Corkum, Kieffer, Legare, and
  Villeneuve}}]{Shiner2011}
\bibinfo{author}{\bibfnamefont{A.~D.} \bibnamefont{Shiner}},
  \bibinfo{author}{\bibfnamefont{B.~E.} \bibnamefont{Schmidt}},
  \bibinfo{author}{\bibfnamefont{C.}~\bibnamefont{Trallero-Herrero}},
  \bibinfo{author}{\bibfnamefont{H.~J.} \bibnamefont{Worner}},
  \bibinfo{author}{\bibfnamefont{S.}~\bibnamefont{Patchkovskii}},
  \bibinfo{author}{\bibfnamefont{P.~B.} \bibnamefont{Corkum}},
  \bibinfo{author}{\bibfnamefont{J.~C.} \bibnamefont{Kieffer}},
  \bibinfo{author}{\bibfnamefont{F.}~\bibnamefont{Legare}}, \bibnamefont{and}
  \bibinfo{author}{\bibfnamefont{D.~M.} \bibnamefont{Villeneuve}},
  \emph{\bibinfo{title}{Probing collective multi-electron dynamics in xenon
  with high-harmonic spectroscopy}}, \bibinfo{journal}{Nat. Phys.}
  \textbf{\bibinfo{volume}{7}}, \bibinfo{pages}{464} (\bibinfo{year}{2011}).

\bibitem[{\citenamefont{Strelkov}(2010)}]{PhysRevLett.104.123901}
\bibinfo{author}{\bibfnamefont{V.}~\bibnamefont{Strelkov}},
  \emph{\bibinfo{title}{Role of autoionizing state in resonant high-order
  harmonic generation and attosecond pulse production}},
  \bibinfo{journal}{Phys. Rev. Lett.} \textbf{\bibinfo{volume}{104}},
  \bibinfo{pages}{123901} (\bibinfo{year}{2010}).
%
\bibitem{Ganeev} R. A. Ganeev, T. Witting, C. Hutchison, F. Frank, M. Tudorovskaya,
M. Lein, W. A. Okell, A. Za\"{\i}r, J. P. Marangos, and J. W. G. Tisch, \textit{Isolated sub-fs XUV pulse generation in Mn
plasma ablation}, Opt. Express \textbf{20}, 25239 (2012).
%

\bibitem[{\citenamefont{Ghimire et~al.}(2011)\citenamefont{Ghimire, DiChiara,
  Sistrunk, Agostini, DiMauro, and Reis}}]{Ghimire2011}
\bibinfo{author}{\bibfnamefont{S.}~\bibnamefont{Ghimire}},
  \bibinfo{author}{\bibfnamefont{A.~D.} \bibnamefont{DiChiara}},
  \bibinfo{author}{\bibfnamefont{E.}~\bibnamefont{Sistrunk}},
  \bibinfo{author}{\bibfnamefont{P.}~\bibnamefont{Agostini}},
  \bibinfo{author}{\bibfnamefont{L.~F.} \bibnamefont{DiMauro}},
  \bibnamefont{and} \bibinfo{author}{\bibfnamefont{D.~A.} \bibnamefont{Reis}},
  \emph{\bibinfo{title}{Observation of high-order harmonic generation in a bulk
  crystal}}, \bibinfo{journal}{Nat Phys} \textbf{\bibinfo{volume}{7}},
  \bibinfo{pages}{138} (\bibinfo{year}{2011}),
  \bibinfo{note}{10.1038/nphys1847}.

\bibitem[{\citenamefont{Vampa et~al.}(2014)\citenamefont{Vampa, McDonald,
  Orlando, Klug, Corkum, and Brabec}}]{PhysRevLett.113.073901}
\bibinfo{author}{\bibfnamefont{G.}~\bibnamefont{Vampa}},
  \bibinfo{author}{\bibfnamefont{C.~R.} \bibnamefont{McDonald}},
  \bibinfo{author}{\bibfnamefont{G.}~\bibnamefont{Orlando}},
  \bibinfo{author}{\bibfnamefont{D.~D.} \bibnamefont{Klug}},
  \bibinfo{author}{\bibfnamefont{P.~B.} \bibnamefont{Corkum}},
  \bibnamefont{and} \bibinfo{author}{\bibfnamefont{T.}~\bibnamefont{Brabec}},
  \emph{\bibinfo{title}{Theoretical analysis of high-harmonic generation in
  solids}}, \bibinfo{journal}{Phys. Rev. Lett.} \textbf{\bibinfo{volume}{113}},
  \bibinfo{pages}{073901} (\bibinfo{year}{2014}).

\end{thebibliography}
\end{document}